\documentclass{KapProc} 
\kluwerbib
\let\lcitebracket(
\let\rcitebracket)
\usepackage[dvips]{graphics,epsfig}

\newcommand{\gtsimeq}{\raisebox{-0.6ex}{$\,\stackrel 
        {\raisebox{-.2ex}{$\textstyle >$}}{\sim}\,$}} 

\begin{document}


\articletitle{Extremely red radio galaxies}
\author{Chris J.\,Willott, Steve Rawlings, Katherine M.\,Blundell} 
\affil{Astrophysics, University of Oxford, UK}
\begin{abstract}
\noindent
 
At least half the radio galaxies at $z > 1$ in the 7C Redshift Survey
have extremely red colours ($R-K > 5$), consistent with stellar
populations which formed at high redshift ($z \gtsimeq 5$). We discuss
the implications of this for the evolution of massive galaxies in
general and for the fraction of near-IR-selected EROs which host AGN,
a result which is now being tested by deep, hard X-ray surveys. The
conclusion is that many massive galaxies undergo at least two active
phases: one at $z \sim 5$ when the black hole and stellar bulge formed
and another at $z \sim 1-2$ when activity is triggered by an event
such as an interaction or merger.

\end{abstract}

\section*{Introduction}

Radio sources are known to reside in massive, luminous host
galaxies. Therefore, they are an excellent way of selecting such
galaxies out to high redshifts and tracing their evolution. The
obscuring torus, which forms the basis of the unified schemes, blocks
the non-stellar nuclear emission from our line-of-sight providing a
much clearer view of the host galaxy properties than is the case for
quasars. 3C radio galaxies at $z \gtsimeq 0.6$ have complex optical
continuum and emission line structures aligned along their radio axes,
which can generally be interpreted as due to recent (jet-induced)
star-formation or non-thermal processes associated with the active
nucleus (e.g. Best et al. 1998). Lacy et al. (1999a) have shown that
the strength of the alignment effect decreases with decreasing radio
luminosity such that fainter samples do not suffer from this problem.

The 7C Redshift Survey (7CRS) is a low-radio frequency (151 MHz)
flux-selected sample in three small patches of sky covering a total
sky area of 0.022 sr. The flux-limit is 0.5 Jy -- a factor of $25
\times$ lower than the revised 3CR sample (Laing et al. 1983). We now
have complete optical/near-infrared identifications for all the radio
sources and $>90$\% spectroscopic redshifts. Further details of the
survey are given in Willott et al. (2001a) and Lacy et
al. (1999b). Only 7 out of 76 sources in the 7C-I and 7C-II regions
lack spectroscopic redshifts. For these sources we have obtained
optical and near-IR photometry in order to constrain their redshifts
using photometric redshift techniques. We find that most of these
seven galaxies have very red colours. A full account of this work is given
in Willott et al. (2001b). A flat cosmology with parameters
$\Omega_{\mathrm M}=0.3$, $\Omega_\Lambda=0.7$ and $H_0=70~ {\rm
km~s^{-1}Mpc^{-1}}$ is assumed throughout.

\section*{Results of photometric redshift fitting}

\begin{figure*}[!ht]
{\hbox to \textwidth{\epsfxsize=0.98\textwidth
\epsfbox{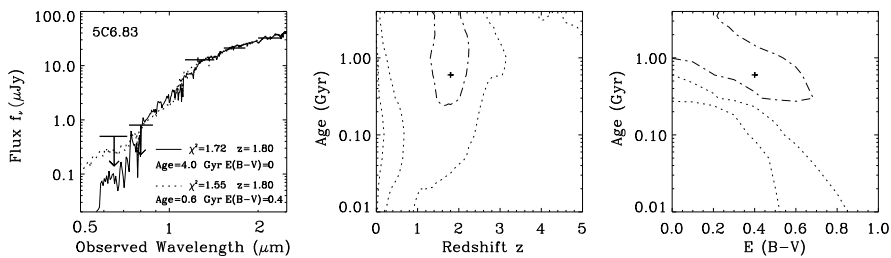}}} 
\vspace{-0.2cm}

{\caption{\label{fig:sed6p83} Example of model-fitting to the optical
to near-IR SEDs of 7CRS radio galaxies. The left panel shows the
broad-band photometric data of 5C6.83 (7CB021111.3+303948; $R>24.5$;
$K=18.3$) with best-fit model galaxies. The best fit unreddened model
is shown as a solid line and the best-fit reddened model as a dotted
line. Details of these fits are given in the bottom-right corner. The
centre and right panels show how the $\chi^2$ of the fit depends upon
redshift, age of the stellar population and reddening. The minimum in
reduced $\chi^2$ (1.55) is shown as a cross.  }} \end{figure*}

$RIJHK$ photometry was obtained for the seven sources without
spectroscopic redshifts. Six of these objects have extremely red
colours ($R-K > 5.5$), similar to those expected from evolved stellar
populations at $z>1$. These data were fit with a range of
instantaneous burst model galaxies (Bruzual \& Charlot in prep.) with
redshift, age and reddening as free parameters. Best-fit models were
found by searching for the minimum in the $\chi^2$ distribution. Fits
with reduced $\chi^2 < 1$ were found for five objects and the
remaining two had best-fit reduced $\chi^2 < 2$. An example of the
SED-fitting is shown in Fig. \ref{fig:sed6p83}.

The best-fit redshifts for these radio galaxies ranges from $z=1.05$
to $z=2.35$ with typical uncertainties $\Delta z = \pm
0.3$. Considering other factors such as the $K-z$ relation and the
lack of emission lines in their optical spectra, we believe that all
the galaxies are likely to fall within the redshift range $1<z<2$. The
flatness of the SEDs from $J$ to $K$ strongly argues against $z>2.5$.
We know from near-IR spectroscopy that the colours are not strongly
affected by emission line contamination, but in two cases there is a
marginal emission line in the near-IR which could be H$\alpha$ at
$z\approx 1.5$, consistent with the SED-fitting. Quite a wide range of
parameters provide acceptable fits for many of the galaxies. The
degeneracy of age and reddening appears to be the strongest cause of
this as shown by the diagonal shape of contours in the age-reddening
plane in Fig. 1 (from old and dust-free at top-left to young and dusty
at bottom-right). In contrast there is little correlation between
redshift and age in the contours and the best-fit redshifts are in
most cases similar for both the reddened and unreddened cases.

\section*{The colours of 7CRS radio galaxies}

\begin{figure*}[!ht]
\vspace{-0.4cm}
{\hbox to \textwidth{\epsfxsize=0.95\textwidth
\epsfbox{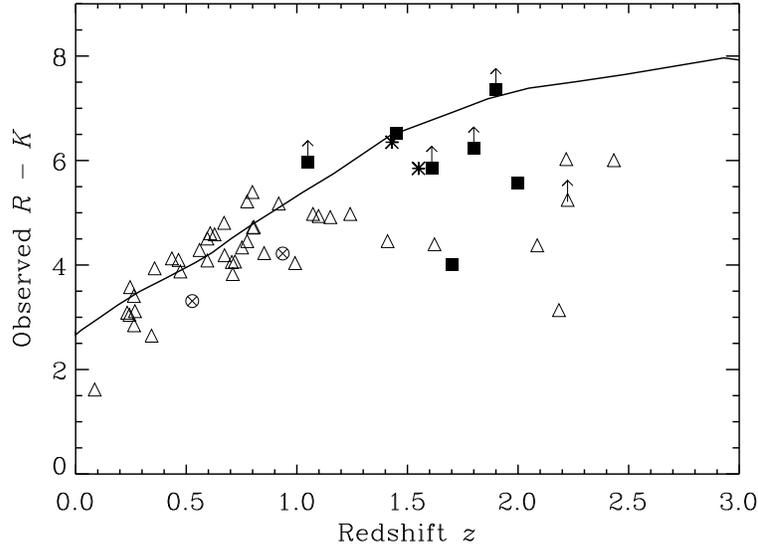}}} 
\vspace{-0.5cm}

{\caption{\label{fig:rkz} Observed $R-K$ colour as a function of
redshift for radio galaxies in the 7C-I and 7C-II regions of the
7CRS. Triangles are narrow-line radio galaxies, crossed circles are
broad-lined radio galaxies and filled squares are the seven 7CRS objects
with photometric redshift estimates. The asterisks are the two red
radio galaxies from the LBDS survey (Dunlop 1999). The solid line
shows the evolution of the expected observed colour of a galaxy which
formed in an instantaneous starburst at redshift $z=5$ using the
models of Bruzual \& Charlot (in prep.). This model provides a good fit
to the upper envelope of the galaxy colours up to at least redshift
$z=2$. }}  \end{figure*}

Since all members of the 7CRS (regions I and II) have been securely
identified, have spectroscopic or photometric redshifts and have $R$
and $K$-band imaging (Willott et al. in prep), we can investigate the
colour evolution of the radio galaxies. In Fig. \ref{fig:rkz} we plot
the observed $R-K$ colour against redshift for the 49 narrow-line
radio galaxies and the 2 broad-line radio galaxies (the 23 quasars are
not plotted because their magnitudes are clearly dominated by
non-stellar emission). The solid curve is a model featuring an
instantaneous starburst at redshift $z=5$. At redshifts $z \leq 1$ the
colours of most of the radio galaxies are close to the model curve
suggesting that these colours result from very old galaxy populations
with little (unobscured) current star-formation. Deviations to bluer
colours can be caused by a small amount of more recent star-formation
or AGN-related processes like scattering of quasar light.

Moving to $z \gtsimeq 1$ we find a marked increase in the scatter of
the $R-K$ colours of the radio galaxies. This is probably due to a
combination of two effects. First, at redshifts higher than $z \sim
0.8$, the observed $R$-band samples the rest-frame light below the
4000 \AA~ break. The ratio of fluxes below and above 4000 \AA~ is a
very strong function of the amount of current or recent
star-formation. Therefore small differences in the amount of recent
star-formation will have a more dramatic effect on the observed colour
at $z > 0.8$.  Lilly \& Longair (1984) showed that the optical/near-IR
colours of 3CR galaxies at $z>1$ are inconsistent with a no-evolution
model. Their observed $R-K \approx 4$ colours are similar to those of
the bluest 7CRS radio galaxies. For the 3C objects these colours are
best explained by recent star-formation in a few cases (Chambers \&
McCarthy 1990), but in most cases, as further evidenced by large
optical polarizations (e.g. di Serego Alighieri et al. 1989), they are
probably caused by an extra non-stellar rest-frame UV component,
typically scattered light from the quasar nucleus (see Best et
al. 1998).

It is clear from Fig. \ref{fig:rkz} that, at $z>1$, the bluer 7CRS
galaxies are much more likely to have redshifts measured from optical
spectroscopy.  This suggests that the bluer sources have stronger
emission lines and are therefore good candidates for having
non-stellar (AGN) and/or starburst components in the rest-frame UV. At
$z > 1$, radio galaxies from the 6CE sample, which has a flux-limit in
between those of 3CRR and the 7CRS, also tend to have strong emission
lines and blue colours (Rawlings et al. 2001). Therefore the reason
that such a high fraction of extremely red radio galaxies at
high-redshift has not been seen in previous complete samples is due to
their higher radio flux-limits selecting only higher luminosity $z>1$
radio galaxies.

\section*{Do the red colours indicate old galaxies?}

We have shown that at least half the $z>1$ radio galaxies in the 7CRS
have red colours ($R-K>5$), consistent with those expected from
evolved stellar populations at these redshifts. In addition, the fact
that the bluer high-redshift radio galaxies have stronger emission
lines, suggests that they too may have underlying host galaxies which
are very red. But can we be certain that these colours are an
indicator of age, rather than the effects of reddening by dust? The
answer is that we cannot be completely certain in the absence of high
signal-to-noise spectroscopy of stellar absorption features. Due to
the faintness of these galaxies in the optical ($I \gtsimeq 24$), this
will have to be performed in the near-infrared utilizing the
low-background available with OH-suppression spectrographs on large
telescopes. Instead, we have to look at circumstantial evidence, such
as that attained for similar objects. Dunlop et al. (1999 and
references therein) have studied two red radio galaxies at $z \sim
1.5$ drawn from a faint radio sample. Deep Keck spectroscopy shows
that the red colours of these galaxies are due to an old stellar
population and not due to reddening by dust. The ages inferred are
still controversial, although ages of $\sim 3$ Gyr remain the best
estimate (Nolan et al. 2001). For these ages observed at $z=1.5$, the
star-formation must have ceased at $z \sim 5$.

\section*{Implications for AGN activity in elliptical galaxies}

Although powerful radio galaxies are very rare objects, we can use our
findings to predict the relationship between other AGN (both
radio-loud and radio-quiet) and the near-infrared selected ERO
population. Extrapolating down the radio luminosity function (Willott
et al. 2001a) to less powerful radio galaxies, we find that if a
similar fraction of these lower power radio galaxies have similar
colours, they imply a space density of red radio galaxies which is
about 3\% of the space density of near-IR selected EROs (Daddi et
al. 2000).  Hence we find that a small, but significant, fraction
($\sim 3$ \%) of field EROs are likely to be hosting radio-loud AGN.

However, it is well-known that luminous radio sources have limited
lifetimes ($\sim 10^8$ years) which are much smaller than the Hubble
time. The time elapsed between $z=2$ and $z=1$ is approximately 3.5
Gyr. Therefore if individual radio sources have lifetimes of only
$\sim 10^8$ years, then the number of galaxies undergoing radio
activity during this period would be a factor of 30 greater than that
observed. Hence all of the near-IR selected EROs could plausibly
undergo such a period of radio activity. A caveat to this is that the
typical lifetimes of weak radio sources such as those which would
dominate the ERO population are not well-constrained and could have
longer lifetimes of $\sim 1$ Gyr.  In such a case, only $\sim 10$\% of
high-$z$ EROs would undergo a period of radio activity at some point
(see Willott et al. 2001b for more details).

The hardness of the X-ray background requires that the space density
of optically-obscured quasars exceeds that of optically-luminous
quasars (e.g. Comastri et al. 1995), which in turn are well known to
outnumber radio-loud quasars by at least an order of magnitude. Many
of the hard X-ray sources discovered in Chandra surveys have very red
galaxy counterparts with weak or absent emission lines (Crawford et
al. 2001; Cowie et al. 2001). These objects are likely to be the
radio-quiet analogues of the 7CRS EROs discussed here. The hard X-ray
properties of the ERO population will be investigated with XMM-Newton
and Chandra surveys of ERO fields. In the HDF-North Caltech area, 4
out of 33 EROs ($R-K>5$) have hard X-ray detections (Hornschemeier et
al 2001). Assuming an elliptical galaxy fraction in the field ERO
population of 70\% (Moriondo et al. 2001; Stiavelli \& Treu 2000) this
corresponds to $\sim 20$\% of these ellipticals being observed to
undergo a phase of AGN activity. Better statistics will accurately
determine the duty cycle of AGN-activity in massive galaxies and
provide constraints on quasar lifetimes (which are currently not
well-constrained for radio-quiet quasars). It seems likely that rather
than being an oddity, AGN activity is common in massive galaxies, not
only during the high-redshift formation epoch, but also at a later
stage once the major episode of star (and black hole) formation has
long since ceased.



\begin{chapthebibliography}{<widest bib entry>}




\bibitem{218} Best P.N., Longair M.S., R\"ottgering H.J.A., 1998,
MNRAS, 295, 549

\bibitem{154} Chambers K.C., McCarthy P.J., 1990, ApJ, 354L, 9

\bibitem{429} Comastri A., Setti G., Zamorani G., Hasinger G., 1995,
A\&A, 296, 1

\bibitem{430} Cowie L.L., et al., 2001, ApJL, in press,
astro-ph/0102306

\bibitem{188} Crawford C.S., Fabian A.C., Gandhi P., Wilman R.J.,
Johnstone R.M., 2001, MNRAS, submitted, astro-ph/0005242

\bibitem{420} Daddi E., et al., 2000, A\&A, 361, 535

\bibitem{145} di Serego Alighieri S., Fosbury R.A.E., Tadhunter C.N.,
Quinn P.J., 1989, Nature, 341, 307

\bibitem{147} Dunlop J.S., 1999, in The Most Distant Radio Galaxies,
ed. P.N. Best, H.J.A. R\"ottgering, M.D. Lehnert, (KNAW Colloq.;
Dordrecht: Kluwer), 14

\bibitem{353} Hornschemeier A.E., et al., 2001, ApJ, in press, astro-ph/0101494

\bibitem{351} Lacy M., Ridgway S.E., Wold M., Lilje P.B., Rawlings S.,
1999a, MNRAS, 307, 420

\bibitem{350} Lacy M., et al., 1999b, MNRAS, 308, 1096

\bibitem{309} Laing R.A., Riley J.M., Longair M.S., 1983, MNRAS, 204,
151

\bibitem{310} Lilly S.J., Longair M.S., 1984, MNRAS, 211, 833

\bibitem{418} Moriondo G., Cimatti A., Daddi E., 2001, A\&A, in press,
astro-ph/0010335

\bibitem{352} Nolan L.A., Dunlop J.S., Jimenez R., Heavens A.F., 2001,
MNRAS, submitted, astro-ph/0103450

\bibitem{355} Rawlings S., Eales S.A., Lacy M., 2001, MNRAS, 322, 523

\bibitem{441} Stiavelli M., Treu T., 2000, To appear in the
proceedings of the conference ``Galaxy Disks and Disk Galaxies'', ASP
Conf. series, eds. Funes and Corsini, astro-ph/0010100

\bibitem{17} Willott C.J., Rawlings S., Blundell K.M., Lacy M., Eales
S.A., 2001a, MNRAS, 322, 536

\bibitem{18} Willott C.J., Rawlings S., Blundell K.M., 2001b, MNRAS,
in press, astro-ph/0011082


\end{chapthebibliography}
\end{document}